\documentstyle[twocolumn,psbox,myletter]{jpsj}

\def\runtitle{Magnetization Plateaus in the Shastry-Sutherland Model}
\def\runauthor{Yoshiyuki {\sc Fukumoto}}

\title{Magnetization Plateaus in the Shastry-Sutherland Model 
for ${\bf SrCu_2(BO_3)_2}$: Results of Fourth-Order Perturbation Expansion 
with a Low-Density Approximation}
\author{Yoshiyuki {\sc Fukumoto}\footnote{E-mail: yfuku@ph.noda.sut.ac.jp} }
\inst{Department of Physics, Faculty of Science and Technology,
Science University of Tokyo,\\ Noda, Chiba 278-8510\\}
\recdate{\hspace{2cm}}

\abst{Magnetization plateaus in the Shastry-Sutherland model for ${\rm SrCu_2(BO_3)_2}$ 
are studied by the perturbation expansion method. The fourth-order effective Hamiltonian 
which describes the dynamics of triplet dimers (TD's) with $S^{\rm tot}_z=1$ in the 
singlet sea is derived and then partially diagonalized for the space that consists of
the TD configurations with the lowest second-order energy. The fourth-order terms are 
treated within a low-density approximation. Our procedure makes clear how TD interactions 
are responsible for the formation of magnetization plateaus. Particularly,
the 1/4-plateau is obtained by the fourth-neighbor TD repulsion in the fourth-order 
perturbation, and a diagonal stripe arrangement of TD's appears at this plateau.}

\kword{${\sf SrCu_2(BO_3)_2}$, Shastry-Sutherland model, magnetization plateau, 
perturbation expansion, Heisenberg antiferromagnet, two dimensions}

\begin{document}
\sloppy
\maketitle
\section{Introduction}
Two-dimensional frustrated Heisenberg antiferromagnets have been considerable interests 
for a long time because of unique magnetic properties which reflect unusual low-energy 
spectra due to a strong geometrical frustration.
A typical example of such frustrated antiferromagnets is the spin-1/2 Kagom\'{e}
antiferromagnet, where a disordered spin-liquid like ground state, a continuum of 
singlet excitations adjacent to the ground state, and gapped triplet excitations 
are obtained by numerical investigations.~\cite{rf:Sindzingre}

\begin{figure}
\begin{center}
\psbox[xsize=6cm]{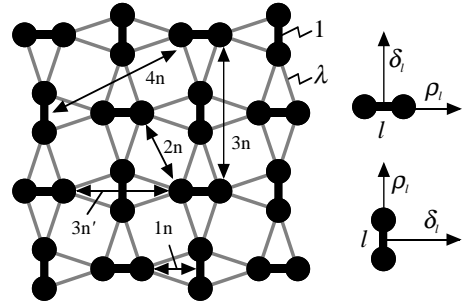}
\end{center}
\caption{Structure of a two-dimensional network formed by 
${\rm Cu}^{2+}$ (closed circles) in ${\rm SrCu_2(BO_3)_2}$. 
The vector $\rho_l$ 
($\delta_l$) is the unit vector parallel (perpendicular) to the 
axis of a dimer at site $l$. The arrows indicate first-neighbor 
(1n), second-neighbor (2n), third-neighbor (3n, 3n') and 
fourth-neighbor (4n) dimer pairs. Note that there exist two 
types of third-neighbor dimer pairs. For brevity, we call that 
indicated by "3n" the third-neighbor pair.}
\label{fig:model}
\end{figure}

For the last two years, there has been a growing interest to a two-dimensional 
frustrated antiferromagnet which is called the Shastry-Sutherland model 
(SSM),~\cite{rf:Shastry} because Kageyame {\it et al.} and Miyahara and Ueda found 
that the SSM is realized in ${\rm SrCu_2(BO_3)_2}$.~\cite{rf:Kageyama1,rf:Miyahara} 
The Hamiltonian of the SSM under the uniform magnetic field $H$ is defined by
\begin{equation}
   {\cal H} = \sum_{\langle i,j\rangle}{\mib S}_{i}\cdot {\mib S}_{j}+
    \lambda\sum_{\langle\!\langle i,j\rangle\!\rangle}{\mib S}_{i}\cdot {\mib S}_{j}
    -MH,
\end{equation}
where the first (second) term represents the intradimer (interdimer) 
coupling, and $M=\sum_i S_i^z$. (See Fig.~\ref{fig:model}.) 
Strength of the interdimer coupling in ${\rm SrCu_2(BO_3)_2}$ is obtained to be 
$\lambda\sim 0.63$.~\cite{rf:Weihong,rf:Miyahara3} A distinctive feature of this model 
is that the orthogonal dimer structure shown in Fig.~\ref{fig:model} leads to the 
exact dimer singlet ground state for $\lambda<\lambda_{\rm c}$, and prohibits one-TD hopping 
up to the fifth order in $\lambda$.~\cite{rf:Shastry, rf:Miyahara}
As for the quantum phase transition at $\lambda=\lambda_{\rm c}$, Koga and Kawakami
obtained $\lambda_{\rm c}=0.68$ and found that the exact dimer singlet ground state 
becomes to be unstable against the plaquette RVB state there.~\cite{rf:Koga}

Low-energy excitation spectra in the SSM have been investigated extensively because
it is expected that the orthogonal dimer structure makes the spectra quite unique:
the localized nature of one TD leads to a narrow band of scattering states of
two TD's, while a TD pair with a short interdimer distance should have inherent 
dynamics in the multiplicity.~\cite{rf:Fukumoto2,rf:Knetter}
For instance, first-neighbor TD pair excitations can 
propagate through the third-order perturbation. As for experimental studies, 
in addition to the spin gap of 3.0 meV, the second spin gap of 4.7 meV was 
observed by ESR and neutron inelastic scattering experiments.~\cite{rf:Nojiri,rf:Kageyama2} 
In particular, the neutron inelastic scattering experiment revealed that the second lowest 
triplet branch shows dispersive behavior although the lowest triplet branch is almost flat. 
In a Raman scattering experiment, a singlet excitation mode at 3.7 meV
was observed.~\cite{rf:Lemmens} These singlet and triplet bound states 
are considered to originate from first-neighbor TD pair excitations. On the other hand, 
first-neighbor TD pair excitations in the quintet sector lead to antibound states,
and the lowest quintet branch consists of second-neighbor TD pair excitations. 
The lowest quintet branch shows rather weak dispersion, and the binding energy is much 
smaller than those of the  singlet and triplet bound states.

The orthogonal dimer structure also gives rise to an unique feature of the magnetization 
process: ${\rm SrCu_2(BO_3)_2}$ shows intermediate magnetization plateaus
at $M=M_{\rm s}/4$ and $M_{\rm s}/8$, where $M_{\rm s}$ represents the saturated
magnetization.~\cite{rf:Kageyama1} To understand these plateaus, several groups 
derived the third-order effective Hamiltonian describing the dynamics of TD with 
$S^{\rm tot}_z=1$ in the singlet sea, and analyzed the resultant effective 
Hamiltonian.~\cite{rf:Miyahara2,rf:Momoi,rf:Fukumoto,rf:Momoi2}
These investigations predicted the appearance of a 1/3-plateau, which was observed in a recent
experiment.~\cite{rf:Onizuka} However, the origin of the 1/4- and 1/8-plateaus
have not been clarified yet. Three scenarios to understand these two plateaus 
have been suggested. The first one is that these plateaus are produced by long-range 
TD repulsions due to higher-order perturbation. The second one
is that some spin interactions omitted in the SSM play an essential role to realize 
these plateaus. For instance, it was shown that a magnetization
plateau at $M=M_{\rm s}/4$ is obtained in a generalized SSM.~\cite{rf:Hartmann} 
The third one is that the relevant cluster to understand these plateaus is not a 
spin dimer, but a spin plaquette, which is based on the fact that ${\rm SrCu_2(BO_3)_2}$ 
locates near the dimer-to-plaquette phase boundary.~\cite{rf:Koga} 

In the present paper, we show that the fourth-order perturbation explains 
the 1/4-plateau invoking no additional spin interactions.
We also describe in detail our previous brief report in Ref.~15 concerned 
with the third-order perturbation.

The present paper is organized as follows: in \S~2, we derive the fourth-order 
effective Hamiltonian with a low-density approximation on the basis of the coupled dimer 
picture of the SSM. The fourth-order effective Hamiltonian is partially diagonalized
for the subspace with the lowest second-order energy, and then the resultant Hamiltonian 
is analyzed by the numerical diagonalization method. In \S~3, we give discussions about our 
calculation. It is particularly discussed how lower-order flip terms affect higher-order TD repulsions. In \S~4, we summarize our results.

\section{Perturbation Expansion}
\subsection{Zeroth-Order Perturbation}
We start with the dimer limit. Eigenstates of a dimer are the singlet state 
$\left|s \right>$ and the triplet states $\left|t_1 \right>$, $\left|t_0 \right>$, 
$\left|t_{-1} \right>$, where the suffix denotes the total $S^z$. The lowest energy 
states with magnetization $M(\geq 0)$ for the whole system consist of $\left|s \right>$ 
and $\left|t_1 \right>$. We assume that the lowest energy states are in the degenerate 
space excluding $\left|t_0 \right>$ and $\left|t_{-1} \right>$ when the interdimer coupling
is taken into account.~\cite{rf:Tachiki} Then we calculate the fourth-order 
effective Hamiltonian
\begin{equation}
\label{eq:Heff}
   {\cal H}_{\rm eff}^{\rm 4PE}=E_{\rm g}+M(\Delta_{\rm sg}^{\rm 4PE}-H)+
                         \sum_{n=1}^{4}\lambda^n V_n,
\end{equation}
where
\begin{equation}
   E_{\rm g}=-\frac{3N_{\rm D}}{4}
\end{equation}
is the absolute ground state energy of the system with $N_{\rm D}$ dimers, and
\begin{equation}
   \Delta_{\rm sg}^{\rm 4PE}=1-\lambda^2-\frac{\lambda^3}{2}-\frac{\lambda^4}{8}
\end{equation}
is the fourth-order series of the spin gap. In eq.~(\ref{eq:Heff}), $V_n$
represents interaction among TD's in the $n$th-order perturbation.
If the TD interaction leads to a kink in the minimum interaction 
energy per dimer, $\epsilon(m)$, as a function of density of TD's, $m(\equiv M/M_{\rm s})$,
then a magnetization plateau is realized. 

To write down the explicit form of $V_n$, it is convenient to introduce a pseudo-spin operator
${\mib I}_l$ for each dimer site. The up and down spin states of ${\mib I}_l$ 
correspond to the original dimer states as follows:
\begin{equation}
   \left|\uparrow \right>_{l}\leftrightarrow \left|t_{1} \right>_{l},\;\;
   \left|\downarrow \right>_{l}\leftrightarrow \left|s \right>_{l}.
\end{equation}
We also define the number operator of up spin and the spin-flip operator:
\begin{equation}
   u_{l}\equiv \frac{1}{2}+I_{l}^{z},
\end{equation}
\begin{equation}
   f_{l,l^{\prime}}\equiv I_{l}^{+}I_{l^{\prime}}^{-}+I_{l}^{-}I_{l^{\prime}}^{+}.
\end{equation}
In terms of the original dimer states, $u_{l}$ means the number operator of TD and 
$f_{l,l^{\prime}}$ the hopping operator of TD.

\subsection{First-Order Perturbation}
\begin{figure}
\begin{center}
\psbox[xsize=7.5cm]{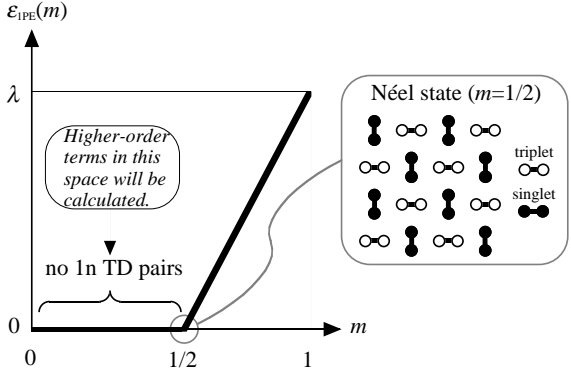}
\end{center}
\caption{First-order approximation of the minimum 
interaction energy $\epsilon_{\rm 1PE}(m)$ 
and TD configuration in the N\'{e}el state.
The dimers with open (closed) circles are in
the triplet (singlet) dimer state. }
\label{fig:em1pe}
\end{figure}

The first-order interaction, $V_1$, is calculated as follows:
\begin{equation}
\label{eq:v1}
   V_1=\frac{1}{2}\sum_{\left<l,l^{\prime}\right>\in {\rm 1n}}
   \hspace{-2mm}u_{l}u_{l^{\prime}},
\end{equation}
which is the nearest-neighbor Ising model on a square lattice. As is well known,
TD's can be arranged on the square lattice avoiding first-neighbor TD pairs for $m\leq 1/2$,
but can not for $m>1/2$. The first-neighbor TD repulsion therefore 
leads to a kink in $\epsilon(m)$ at $m=1/2$ as shown in Fig.~\ref{fig:em1pe}, which 
means the appearance of 1/2-plateau. The N\'{e}el type arrangement of TD's is stabilized 
at the 1/2-plateau. The lower critical field of the 1/2-plateau is given by
$\Delta_{\rm sg}^{\rm 1PE}(=1)$. The most important fact is that all the configurations 
without first-neighbor TD pairs for $0\leq m\leq 1/2$ degenerate at the critical field.
So when another TD repulsion is taken into account, a new magnetization plateau is obtained
even if the strength of the repulsion is small.

We are going to use the same argument to study if higher-order TD repulsions 
lead to additional plateaus. The situation becomes more and more complicated
as the order of the perturbation becomes higher: in ${\cal H}_{\rm eff}^{\rm 4PE}$,
there exist many types of diagonal interactions and off-diagonal interactions.
To overcome this difficulty, we utilize the fact of $V_1$ being a diagonal operator,
which enables us to carry out partial diagonalization of ${\cal H}_{\rm eff}^{\rm 4PE}$.
To be precise, we direct our attention to the subspace without first-neighbor TD pairs for 
$m\leq 1/2$, in which the first-order energy is the lowest value.
We write the new effective Hamiltonian as follows:
\begin{equation}
\label{eq:barHeff}
   \bar{\cal H}_{\rm eff}^{\rm 4PE}=E_{\rm g}+M(\Delta_{\rm sg}^{\rm 4PE}-H)+
   \sum_{n=2}^4\lambda^n\bar{V}_{n}\;\;\;\mbox{for $M\leq \frac{M_{\rm s}}{2}$,}
\end{equation}
where the constraint excluding first-neighbor TD pairs
\begin{equation}
   \bar{u}_{l}\bar{u}_{l\pm\rho_l}=0\;\;\mbox{for any $l$}
\end{equation}
is imposed. The bar on operators means that the operators are defined
in the constrained space.

\subsection{Second-Order Perturbation}
The second-order interaction, $V_2$, is calculated as follows:
\begin{eqnarray}
\label{eq:v2}
   V_{2}&=&
   \frac{1}{2}\hspace{-1mm}\sum_{\left<l,l^{\prime}\right>\in {\rm 1n}}
   \hspace{-2mm}u_{l}u_{l^{\prime}}
   +\frac{1}{2}\hspace{-1mm}\sum_{\left<l,l^{\prime}\right>\in {\rm 3n}}
   \hspace{-2mm}u_{l}u_{l^{\prime}}
   -\frac{1}{2}\sum_{l}u_l u_{l+\rho_l} u_{l-\rho_l}\nonumber\\
   &+&\frac{1}{4}(F_{\rm 1n\rightarrow 2n}+F_{\rm 2n\rightarrow 1n}+
   F_{\rm 1n\leftrightarrow 1n}).
\end{eqnarray}
Here,
\begin{equation}
   F_{\rm 1n\rightarrow 2n}=
   \sum_{l}(u_{l+\rho_l}-u_{l-\rho_l})I_l^{-}(I_{l+\delta_l}^{+}-I_{l-\delta_l}^{+}),
\end{equation}
\begin{equation}
   F_{\rm 2n\rightarrow 1n}=F_{\rm 1n\rightarrow 2n}^{\dagger},
\end{equation}
and
\begin{equation}
   F_{\rm 1n\leftrightarrow 1n}=\sum_{l}u_{l}f_{l-\delta_l,l+\delta_l},
\end{equation}
are correlated hopping terms: $F_{\rm 1n\rightarrow 2n}$ ($F_{\rm 2n\rightarrow 1n}$) 
annihilates a first (second) neighbor TD pair and creates a second (first) neighbor
TD pair, and $F_{\rm 1n\leftrightarrow 1n}$ annihilates a first-neighbor TD pair and 
creates another first-neighbor TD pair. 
As we will see below, $F_{\rm 1n\rightarrow 2n}$ and $F_{\rm 2n\rightarrow 1n}$ play 
a crucial role to produce an intermediate plateau at $m=1/4$. 

As mentioned previously, we calculate $\bar{V}_{2}$ in the subspace without
first-neighbor TD pairs. Noting that the matrix elements of the three-body and 
correlated hopping terms are zero if initial and/or final states contain no first-neighbor 
TD pairs, we get
\begin{equation}
\label{eq:barv2}
   \bar{V}_{2}=\frac{1}{2}\hspace{-1mm}\sum_{\left<l,l^{\prime}\right>\in {\rm 3n}}
   \hspace{-2mm}\bar{u}_{l}\bar{u}_{l^{\prime}}.
\end{equation}
This equation shows that the third-neighbor TD repulsion dominates the low-energy physics
in the second-order perturbation.

\begin{figure}
\begin{center}
\psbox[xsize=6.5cm]{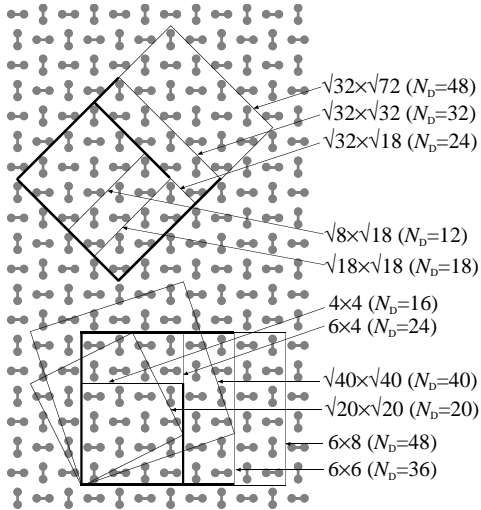}
\end{center}
\caption{Shapes of finite size clusters
used in the numerical studies.}
\label{fig:boundary}
\end{figure}

\begin{figure}
\begin{center}
\psbox[xsize=11cm]{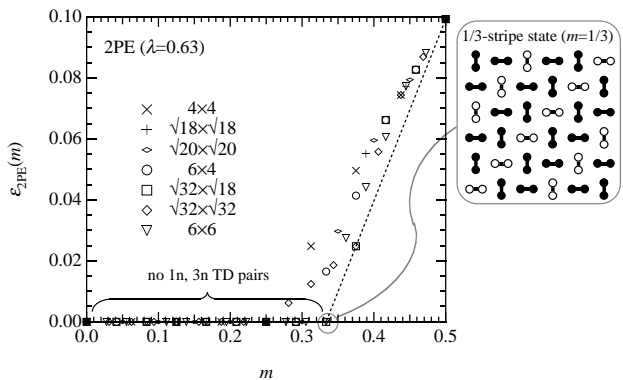}
\end{center}
\caption{Second-order results of minimum 
interaction energy, $\epsilon_{\rm 2PE}(m)$, 
at $\lambda=0.63$, and TD configuration 
in the 1/3-stripe state.}
\label{fig:stripe}
\end{figure}

The third-neighbor TD repulsion in $\bar{V}_{2}$ increases the energy of the N\'{e}el state,
which is the lowest energy state at $m=1/2$, by $\lambda^2 N_{\rm D}/4$. 
Thus the second-order perturbation leads to a new kink in $\epsilon(m)$.
To determine the position of the kink, we calculate the second-order approximation of 
the minimum interaction energy, $\epsilon_{\rm 2PE}(m)$, for various finite size clusters
up to $N_{\rm D}=36$. The shapes of clusters used in the present work are given in 
Fig.~\ref{fig:boundary}, where the periodic boundary condition is imposed. 
We show the result of $\epsilon_{\rm 2PE}(m)$ in Fig.~\ref{fig:stripe}. 
It is found that the kink appears at $m=1/3$. The lowest-energy state at $m=1/3$ is uniquely 
determined as the 1/3-stripe state which is defined in Fig.~\ref{fig:stripe}. 
This result is the same as the results obtained by Miyahara and 
Ueda~\cite{rf:Miyahara2} and Momoi and Totsuka.~\cite{rf:Momoi,rf:Momoi2}


Equation (\ref{eq:barv2}) shows that no off-diagonal terms exist in the second-order 
interaction, which means that higher order terms may play an important role.
Thus we can carry out partial diagonalization again.
Then we study the subspace without first- and third-neighbor TD pairs for $m\leq 1/3$, 
which is the lowest energy states within the second-order approximation. 
We write this effective Hamiltonian in the following form:
\begin{equation}
   \tilde{{\cal H}}_{\rm eff}^{\rm 4PE}=E_{\rm g}+M(\Delta_{\rm sg}^{\rm 4PE}-H)+
      \sum_{n=3}^{4}\lambda^n\tilde{V}_{n}\;\;\mbox{for $M\leq\frac{M_{\rm s}}{3}$,}
\end{equation}
with the constraint
\begin{equation}
   \tilde{u}_{l}\tilde{u}_{l\pm\rho_l}=\tilde{u}_{l+\rho_l}\tilde{u}_{l-\rho_l}=0\;\;
   \mbox{for any $l$}.
\end{equation}
The tilde on operators indicates the exclusion of first- and 
third-neighbor TD pairs.

\begin{figure}
\begin{center}
\psbox[xsize=6cm]{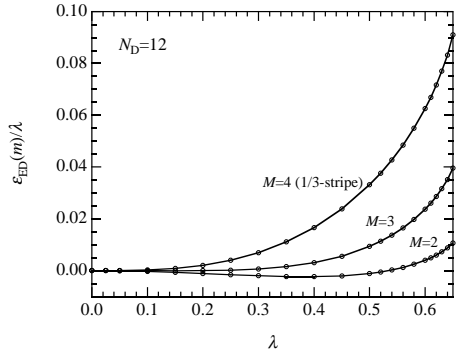}
\end{center}
\caption{Exact diagonalization results
of the minimum interaction energy, 
$\epsilon_{\rm ED}(m)$, for the
12-dimer cluster.}
\label{fig:em(ed)}
\end{figure}

As long as the interdimer coupling $\lambda$ is small enough, the lowest energy states 
for $m\leq 1/3$ are in the subspace without first- and third-neighbor TD pairs due to
TD repulsions up to the second-order perturbation. On the other hand, when $\lambda$ is 
increased, the energy-gain due to TD hopping becomes to be more important than the TD 
repulsions. There should be an energy level crossing between these two regions.
Our assumption for the restriction of the subspace is expected to be valid for wider 
range of $\lambda$ as the density of TD's is lower, because the TD hopping in low-order 
perturbation is achieved by the correlated hopping. To check the validity of our perturbation 
theory for ${\rm SrCu_2(BO_3)_2}$ with $\lambda\sim 0.63$, we use the exact diagonalization 
method and calculate the minimum interaction energy, $\epsilon_{\rm ED}(m)$, on the 12-dimer 
cluster which matches the structure of the N\'eel and 1/3-stripe states. The result is shown 
in Fig.~\ref{fig:em(ed)}. We find no energy level crossing up to $\lambda\sim 0.63$, 
which means that the lowest energy states in ${\rm SrCu_2(BO_3)_2}$ are adiabatically 
connected to the states retained in our perturbation theory. This gives a support to 
the validity of our approach to ${\rm SrCu_2(BO_3)_2}$.

\subsection{Third-Order Perturbation}
We turn to the third-order interaction, $V_3$. The calculation of this term
is somewhat tedious, but can be achieved straightforwardly. The result is given by
\begin{eqnarray}
\label{eq:v3}
   V_{3}=&&
   -\frac{1}{8}\hspace{-1mm}\sum_{\left<l,l^{\prime}\right>\in {\rm 1n}}
   \hspace{-2mm}u_{l}u_{l^{\prime}}
   +\frac{1}{4}\hspace{-1mm}\sum_{\left<l,l^{\prime}\right>\in {\rm 2n}}
   \hspace{-2mm}u_{l}u_{l^{\prime}}
   +\frac{3}{4}\hspace{-1mm}\sum_{\left<l,l^{\prime}\right>\in {\rm 3n}}
   \hspace{-2mm}u_{l}u_{l^{\prime}}
   \nonumber \\ &&
   +\frac{3}{8}F_{\rm 1n\rightarrow 2n}+\frac{1}{4}F_{\rm 2n\rightarrow 1n}
   +\frac{3}{8}F_{\rm 1n\leftrightarrow 1n}
   \nonumber \\ &&
   +(\mbox{three-body terms}).
\end{eqnarray}
We find in $V_{3}$ that second-neighbor TD repulsion appears for the first time in 
the third order. We also note that the coefficient of $F_{\rm 1n\rightarrow 2n}$ is 
different from that of $F_{\rm 2n\rightarrow 1n}$, i.e., $V_{3}$ is not an Hermite operator.
All the off-diagonal terms in $V_{3}$ take the form of correlated hopping.
The explicit expressions of the three-body terms are given in Ref.~16.
The three-body terms, however, play no role in our treatment, because there are no 
matrix elements of the three-body terms in the subspace without first-neighbor 
TD pairs and we will use the two-body approximation in the fourth-order perturbation.

\begin{figure}
\begin{center}
\psbox[xsize=6.5cm]{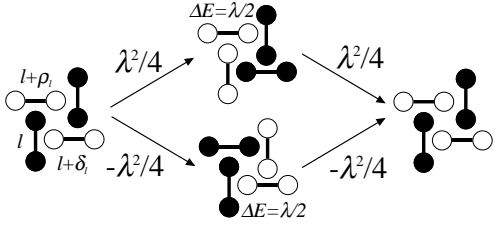}
\end{center}
\caption{Second-order processes which cancel
the second-neighbor TD repulsion in $V_3$. The number
attached on each arrow represents the matrix
element.}
\label{fig:3PEattraction}
\end{figure}

Using eqs.~(\ref{eq:v1}), (\ref{eq:v2}) and (\ref{eq:v3}), we obtain
\begin{equation}
   \tilde{V}_{3}=\frac{1}{8}\sum_{l}(\tilde{u}_{l+\rho_l}+\tilde{u}_{l+\rho_l})
      (\tilde{u}_{l+\delta_l}\tilde{u}_{l-\delta_l}+\tilde{f}_{l+\delta_l,l-\delta_l}).
\end{equation}
Note that the second-neighbor repulsion exists in $V_3$, but does not in $\tilde{V}_{3}$.
The processes which cancel the second-neighbor TD repulsion are shown in 
Fig.~\ref{fig:3PEattraction}. We assume there exist two TD's at 
$l+\rho_l$ and $l+\delta_l$ in the initial state.
Operating $(\lambda^2/4)F_{\rm 2n\rightarrow 1n}$ on the initial state, we get the intermediate
states with a first-neighbor TD pair, whose energy is higher by $\Delta E=\lambda/2$ than 
the initial state. Operating $(\lambda^2/4)F_{\rm 1n\rightarrow 2n}$ on the intermediate 
states, we obtain the final state which is the same as the initial state. These second-order 
processes give a second-neighbor TD attraction 
$-(\lambda^3/4)\bar{u}_{l+\rho_l}\bar{u}_{l+\delta_l}$, 
which cancels the second-neighbor TD repulsion in the first-order process.

\begin{figure}
\begin{center}
\psbox[xsize=6.5cm]{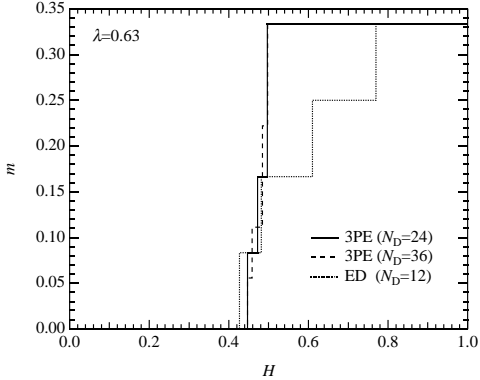}
\end{center}
\caption{Magnetization curve at $\lambda=0.63$.
The solid and long dashed lines are the results
of the third-order perturbation (3PE), and the short
dashed line is the result of the exact diagonalization 
(ED) of the SSM for the 12-dimer cluster.}
\label{fig:mh(3pe)}
\end{figure}

The third-order diagonal term in $\tilde{V}_{3}$, which is a three-body repulsion, 
does not increase the energy of the 1/3-stripe state. Thus we obtain no plateau originating 
in the third-order effect. From this order, there exist correlated hopping terms.
To study the effects of the correlated hopping terms on the magnetization
curve, we diagonalize numerically $\tilde{V}_{3}$ on finite size clusters with 
$N_{\rm D}=24,\;36$. The resultant magnetization curve at $\lambda=0.63$ is
shown in Fig.~\ref{fig:mh(3pe)} together with the exact magnetization curve for the cluster 
with $N_{\rm D}=12$. The correlated hopping terms result in a finite width of
the magnetic field region between $m=0$ and 1/3. However, this region is much narrower 
than that in the exact result. This result indicates that the energy-gain due to the
third-order correlated hopping is rather small. So we expect that a magnetization plateau
appears if there exists a TD repulsion responsible for a magnetization plateau in 
the next order.

\subsection{Fourth-Order Perturbation}
Up to now, all the perturbation processes have been taken into account in our calculation.
In the fourth-order perturbation, such calculation is far from transparent. 
So we invoke the fact that we are now interested in low density region of TD's, $m\leq 1/3$,
and neglect three or more body terms in $\tilde{V}_4$.

The fourth-order term $V_4$ takes the following form:
\begin{eqnarray}
  V_4=&&
  -\frac{9}{16}\hspace{-1mm}\sum_{\left<l,l^{\prime}\right>\in {\rm 1n}}
  \hspace{-2mm}u_{l}u_{l^{\prime}}
  +\frac{3}{8}\hspace{-1mm}\sum_{\left<l,l^{\prime}\right>\in {\rm 2n}}
  \hspace{-2mm}u_{l}u_{l^{\prime}}
  +\frac{1}{8}\hspace{-1mm}\sum_{\left<l,l^{\prime}\right>\in {\rm 3n}}
  \hspace{-2mm}u_{l}u_{l^{\prime}}
  \nonumber \\ &&
  +\frac{1}{8}\hspace{-1mm}\sum_{\left<l,l^{\prime}\right>\in {\rm 4n}}
  \hspace{-2mm}u_{l}u_{l^{\prime}}
  -\frac{1}{32}F_{{\rm 1n}\rightarrow {\rm 2n}}
  +\frac{9}{32}F_{{\rm 2n}\rightarrow {\rm 1n}}
  +\frac{1}{8}F_{{\rm 1n}\leftrightarrow {\rm 1n}}
  \nonumber \\ &&
  +\frac{1}{16}\sum_{l}[(u_{l+2\rho_l+\delta_l}+u_{l+\delta_l}
     -u_{l+2\rho_l-\delta_l}-u_{l-\delta_l})
  \nonumber \\ && \hspace*{1.5cm}
     \times f_{l,l+\rho_l}-(\rho_l \rightarrow -\rho_l)]
  \nonumber \\ &&
  +\frac{1}{32}\sum_{l}[(u_{l-\rho_l}+u_{l+2\rho_l+\delta_l}
     -u_{l+\rho_l}-u_{l+\delta_l})
  \nonumber \\ && \hspace*{1.5cm}
     \times f_{l,l+\rho_l+\delta_l}+(\delta_l \rightarrow -\delta_l)]
  \nonumber \\ &&
  +\frac{1}{16}\sum_{l}[u_{l+\rho_l}(f_{l,l+2\rho_l+\delta_l}-f_{l,l+2\rho_l-\delta_l})
     -(\rho_l \rightarrow -\rho_l)]
  \nonumber \\ &&
  -\frac{1}{4}\sum_{l}(I^{+}_{l}I^{+}_{l+\rho_l+\delta_l}
                       I^{-}_{l+\rho_l}I^{-}_{l+\delta_l}+{\rm h.c.})
  \nonumber \\ &&
  +\frac{1}{32}\sum_{l}f_{l,l+\rho_l+\delta_l}f_{l+\rho_l,l+\delta_l}
  \nonumber \\ &&
  +(\mbox{three or more body terms}).
\end{eqnarray}
We find in $V_4$ that fourth-neighbor TD repulsion appears for the first time 
in the fourth order.
The matrix element of the second-neighbor TD repulsion is three times as strong as that of
the fourth-neighbor repulsion. As for the off-diagonal part, we find that TD pair 
hopping is possible from this order.

\begin{figure}
\begin{center}
\psbox[xsize=7.0cm]{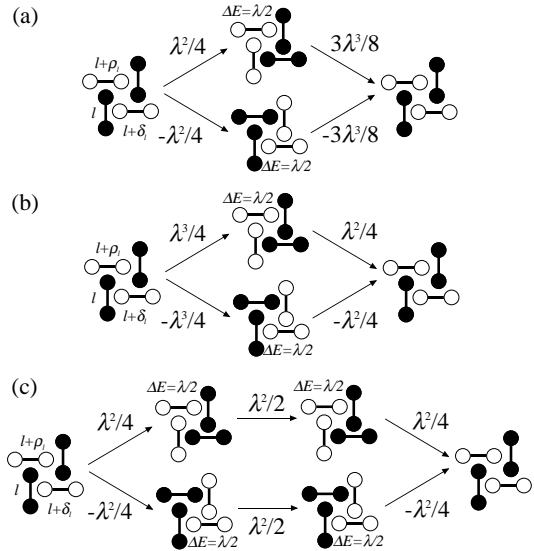}
\end{center}
\caption{Second- and third-order processes
which cancel the second-neighbor TD repulsion in $V_4$.}
\label{fig:4PEattraction}
\end{figure}

Calculation of matrix elements in the two-body approximation of $\tilde{V}_4$ is  easily
carried out by introducing the two-TD basis set in the momentum space,
because it is reduced to partial diagonalization of a finite size matrix
due to luck of the one-TD hopping.
The result is given by
\begin{eqnarray}
  \tilde{V}_{4}&\simeq&\;\;
   \frac{1}{8}\sum_{\left<l,l^{\prime}\right>\in {\rm 4n}}
     \tilde{u}_{l}\tilde{u}_{l^{\prime}}
  \nonumber\\&&
  -\frac{1}{16}\sum_{l}[\tilde{f}_{l,l+\delta_l}
     (\tilde{u}_{l+\rho_l-\delta_l}-\tilde{u}_{l-\rho_l-\delta_l})
     -(\delta_l \rightarrow -\delta_l)]
  \nonumber\\&&
  +\frac{3}{16}\sum_{l}\tilde{f}_{l,l+2\rho_l}
     (\tilde{u}_{l+\rho_l+\delta_l}+\tilde{u}_{l+\rho_l-\delta_l})
  \nonumber\\&&
  -\frac{1}{16}\sum_{l}\tilde{f}_{l,l+2\delta_l}
     (\tilde{u}_{l+\rho_l+\delta_l}+\tilde{u}_{l-\rho_l+\delta_l})
  \nonumber\\&&
  +\frac{1}{16}\sum_{l}[\tilde{f}_{l,l+2\rho_l+2\delta_l}\tilde{u}_{l+\rho_l+2\delta_l}
     +(\delta_l \rightarrow -\delta_l)]
  \nonumber\\&&
  -\frac{1}{4}\sum_{l}(\tilde{I}^{+}_{l}\tilde{I}^{+}_{l+\rho_l+\delta_l}
                       \tilde{I}^{-}_{l+\rho_l}\tilde{I}^{-}_{l+\delta_l}+{\rm h.c.}).
\end{eqnarray}
Note that the second-neighbor TD repulsion exists in $V_4$, but does not in $\tilde{V}_{4}$.
The processes which cancel the second-neighbor TD repulsion are shown in 
Fig.~\ref{fig:4PEattraction}.
In Fig.~\ref{fig:4PEattraction}(a), we use $(\lambda^2/4)F_{\rm 2n\rightarrow 1n}$
to obtain the intermediate states, and  $(3\lambda^3/8)F_{\rm 1n\rightarrow 2n}$
to obtain the final state. In Fig.~\ref{fig:4PEattraction}(b), we use 
$(\lambda^3/4)F_{\rm 2n\rightarrow 1n}$ to obtain the intermediate states, and 
$(\lambda^2/4)F_{\rm 1n\rightarrow 2n}$ to obtain the final state.
These second-order processes (a) and (b) give second-neighbor TD attractions
$-(3\lambda^4/8)\bar{u}_{l+\rho_l}\bar{u}_{l+\delta_l}$ and 
$-(\lambda^4/4)\bar{u}_{l+\rho_l}\bar{u}_{l+\delta_l}$, respectively.
The process in Fig.~\ref{fig:4PEattraction}(c) is a third-order one, where
$(\lambda^2/4)F_{\rm 2n\rightarrow 1n}$ is used to obtain the intermediate states
and $(\lambda^2/4)F_{\rm 1n\rightarrow 2n}$ is used to obtain the final state.
This process gives a second-neighbor TD repulsion 
$(\lambda^4/4)\bar{u}_{l+\rho_l}\bar{u}_{l+\delta_l}$.
Summing up the contributions of these three, we obtain a second-neighbor TD attraction of 
$-(3\lambda^4/8)\bar{u}_{l+\rho_l}\bar{u}_{l+\delta_l}$, which 
cancels the second-neighbor TD repulsion in the first-order process.

\begin{figure}
\begin{center}
\psbox[xsize=11cm]{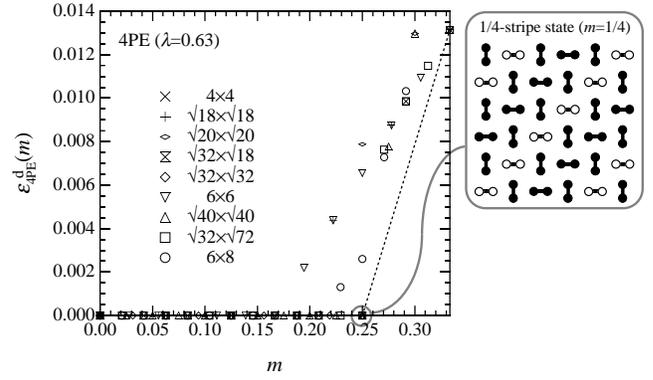}
\end{center}
\caption{Fourth-order results of minimum 
interaction energy $\epsilon_{\rm 4PE}^{\rm d}(m)$,
for the diagonal and low-density approximation
of $\tilde{{\cal H}}_{\rm eff}^{\rm 4PE}$,
and TD configuration in the
1/4-stripe state, where we use $\lambda=0.63$.}
\label{fig:stripeB}
\end{figure}

In $\tilde{V}_{4}$, there exists the fourth-neighbor TD repulsion which increases the 
energy of the 1/3-stripe state by $\lambda^4 N_{\rm D}/12$. The fourth-neighbor TD 
repulsion therefore leads to a kink in $\epsilon(m)$. 
To determine the position of the kink, we neglect TD hopping terms in 
$\tilde{{\cal H}}_{\rm eff}^{\rm 4PE}$ and calculate the diagonal approximation of 
the interaction energy, $\epsilon_{\rm 4PE}^{\rm d}(m)$, on various finite-size clusters 
up to $N_{\rm D}=48$. The result is shown in Fig.~\ref{fig:stripeB}. It is found that 
there is the kink at $m=1/4$. The lowest-energy state at $m=1/4$ is uniquely 
determined as the 1/4-stripe state which is defined in Fig.~\ref{fig:stripeB}. 


\begin{figure}
\begin{center}
\psbox[xsize=6.5cm]{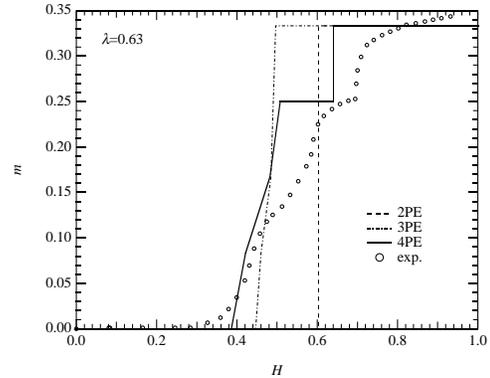}
\end{center}
\caption{Magnetization curve at $\lambda=0.63$
calculated by using the perturbation expansion
method up to the second order (2PE), the third order 
(3PE) and the fourth order (4PE). 
Calculations of the third- and fourth-order 
perturbation are carried out for the $\sqrt{32}\times\sqrt{18}$ 
cluster ($N_{\rm D}=24$). The circles represent 
experimental data obtained by Onizuka {\it et al.},~\cite{rf:Onizuka}
where we use the intradimer coupling 84 K and the
$g$-value 2.05.}
\label{fig:mh}
\end{figure}

We calculate the magnetization curve of $\tilde{{\cal H}}_{\rm eff}^{\rm 4PE}$ on the 
$\sqrt{32}\times\sqrt{18}$ cluster. The results are shown in Fig.~\ref{fig:mh}
together with the second- and third-order results and the experimental result
obtained by Onizuka {\it et al.}~\cite{rf:Onizuka} In this figure, midpoints of each 
step due to hopping terms are connected by straight line to distinguish these from steps 
due to TD repulsions. Our theoretical result explains the 1/4-plateau observed 
experimentally. The width of the 1/4-plateau is also consistent with that of 
the experiment.

\section{Discussions}
\begin{figure}
\begin{center}
\psbox[xsize=7.5cm]{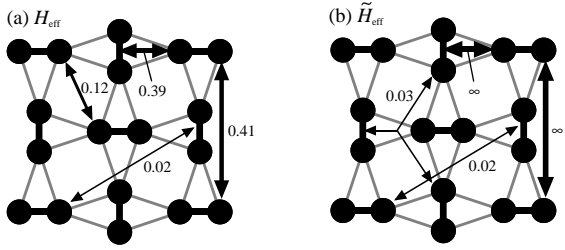}
\end{center}
\caption{Repulsions among TD's 
in (a) ${\cal H}_{\rm eff}^{\rm 4PE}$ and (b) 
$\tilde{{\cal H}}_{\rm eff}^{\rm 4PE}$.
The number attached on each arrows represents 
the matrix element at $\lambda=0.63$.}
\label{fig:repulsion}
\end{figure}

In our perturbation theory, we first calculate ${\cal H}_{\rm eff}^{\rm 4PE}$, and then 
calculate $\tilde{{\cal H}}_{\rm eff}^{\rm 4PE}$ for $m\leq 1/3$ in the subspace without
first- and third-neighbor TD pairs. We here comment on the TD repulsions in these two 
Hamiltonians. In Fig.~\ref{fig:repulsion}(a), we show the two-body repulsions in 
${\cal H}_{\rm eff}^{\rm 4PE}$. There exist first-, third-, second- and 
fourth-neighbor TD repulsions in ${\cal H}_{\rm eff}^{\rm 4PE}$, which appear in the 
first-, second-, third- and fourth-order perturbation for the first time, respectively. 
At $\lambda=0.63$, the matrix elements of these repulsions are given by 0.39, 0.41, 0.12 
and 0.02 for first-, third-, second- and fourth-neighbor TD pairs.
In Fig.~\ref{fig:repulsion}(b), we show all the repulsions in the low-density
approximation of $\tilde{{\cal H}}_{\rm eff}^{\rm 4PE}$.
Here the first- and third-neighbor repulsions are treated as the constraint, 
which means that the matrix elements of these repulsions are approximated as the infinity.
The other repulsions are the fourth-neighbor repulsion and the three-body repulsion. 

\begin{figure}
\begin{center}
\psbox[xsize=11cm]{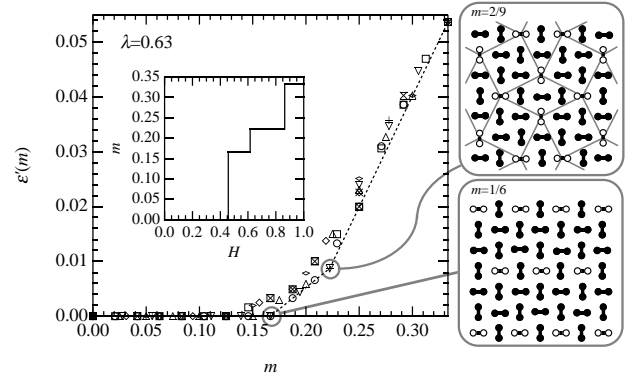}
\end{center}
\caption{Interaction energy, $\epsilon^{\prime}(m)$,
and the lowest energy configurations at $m=2/9$ and 
1/6 for $\tilde{\cal{H}}_{\rm eff}^{\prime}$, where 
we use $\lambda=0.63$. The correspondence of 
the symbols to the clusters is the same as that in 
Fig.~\ref{fig:stripeB}. The inset shows the 
magnetization curve. In the configuration at 
$m=2/9$, the interaction energy originates from 
the fourth-neighbor repulsion indicated by gray
lines.}
\label{fig:mhp}
\end{figure}

The vanishing of the second-neighbor TD repulsion in $\tilde{{\cal H}}_{\rm eff}^{\rm 4PE}$
is essential to obtain the 1/4-plateau. To illustrate this, we consider the following 
Hamiltonian:
\begin{eqnarray}
  \tilde{\cal{H}}_{\rm eff}^{\prime}&=&E_{\rm g}+M(\Delta_{\rm sg}^{\rm 4PE}-H)
  +\frac{\lambda^3(2+3\lambda)}{8}\sum_{\left<l,l^{\prime}\right>\in {\rm 2n}}
     \tilde{u}_{l}\tilde{u}_{l^{\prime}}
  \nonumber\\&&
  +\frac{\lambda^4}{8}\sum_{\left<l,l^{\prime}\right>\in {\rm 4n}}
     \tilde{u}_{l}\tilde{u}_{l^{\prime}},
\end{eqnarray}
which is obtained by omitting the three-body repulsions and the second-neighbor TD attractions
which originate from second- or third-order processes, and all flip terms from the low 
density approximation of $\tilde{{\cal H}}^{\rm 4PE}_{\rm eff}$. Note that the second-neighbor
TD repulsion is much stronger than the fourth-neighbor TD repulsion.
We carry out the finite-size study of $\tilde{{\cal H}}^{\prime}_{\rm eff}$
using the clusters up to $N_{\rm D}=48$. In Fig.~\ref{fig:mhp}, we show 
the interaction energy, magnetization curve, and dimer configurations at plateaus.
We find that there exist two plateaus at $m=2/9$ and 1/6,
which contradict to the experimental results. 
It is obvious that the second-neighbor TD repulsion prevents the 1/4-plateau
with the diagonal stripe TD arrangement.
Thus, the 1/4-plateau is never obtained if one fails to notice the cancellation of 
the second-neighbor repulsion.~\cite{rf:Fukumoto3}

It should be mentioned that the existence of the second-neighbor repulsion in 
${\cal H}^{\rm 4PE}_{\rm eff}$ from the third order has been known, and
the possibility of the 1/4-plateau with the diagonal stripe TD arrangement has
been pointed out.~\cite{rf:Miyahara2} However, it has remained as an open problem
how this 1/4-plateau is stabilized although the second-neighbor repulsion exists.
This question is resolved by the present results of the cancellation of the second-neighbor 
repulsion due to the correlated hopping terms of $F_{{\rm 1n}\rightarrow {\rm 2n}}$ 
and $F_{{\rm 2n}\rightarrow {\rm 1n}}$, and the existence of the fourth-neighbor repulsion 
in the fourth-order perturbation.

\section{Summary}
We have been studied magnetization plateaus in the Shastry-Sutherland model. 
We have derived the fourth-order effective Hamiltonian, ${\cal H}^{\rm 4PE}_{\rm eff}$, 
and then we have carried out partial diagonalization to get 
$\tilde{{\cal H}}^{\rm 4PE}_{\rm eff}$ for the subspace
without first- and third-neighbor triplet dimer pairs, in which second-order energy takes 
the lowest value. We have checked the validity of our approach to ${\rm SrCu_2(BO_3)_2}$
by using the exact diagonalization method for the 12-dimer cluster. 
It has been shown that in the subspace excluding first-neighbor TD pairs
the correlated hopping terms, $F_{{\rm 1n}\rightarrow {\rm 2n}}$ and 
$F_{{\rm 2n}\rightarrow {\rm 1n}}$, lead to the cancellation of
the second-neighbor repulsion in ${\cal H}^{\rm 4PE}_{\rm eff}$.
Then the fourth-neighbor repulsion is responsible for 
the 1/4-plateau observed experimentally. At the 1/4-plateau,
the diagonal stripe arrangement of TD's is stabilized.

\acknowledgements
The author would like to thank Professor A. Oguchi for useful discussions and critical 
reading of the manuscript. We have used a part of the codes provided by H. Nishimori in 
TITPACK Ver.~2.


\begin{thebibliography}{99}
\bibitem{rf:Sindzingre} P. Sindzingre, G. Misguich, C. Lhuillier, B. Bernu, L. Pierre, 
   Ch. Waldtmann, H. U-. Everts: Phys. Rev. Lett. {\bf 84} (2000) 2953.
\bibitem{rf:Shastry} S. Shastry and B. Sutherland: Physica {\bf 108}B (1981) 1069.
\bibitem{rf:Kageyama1} H. Kageyama, K. Yoshimura, R. Stern, N. V. Mushnikov, K. Onizuka, 
   M. Kato, K. Kosuge, C. P. Slichter, T. Goto and Y. Ueda:
   Phys. Rev. Lett. {\bf 82} (1999) 3168.
\bibitem{rf:Miyahara} S. Miyahara and K. Ueda: Phys. Rev. Lett. {\bf 82} (1999) 3701.
\bibitem{rf:Weihong} Z. Weihong, C. J. Hamer and J. Oitmaa: Phys. Rev. B{\bf 60} (1999) 6608.
\bibitem{rf:Miyahara3} S. Miyahara and K. Ueda: cond-mat/0004260.
\bibitem{rf:Koga} A. Koga and N. Kawakami: Phys. Rev. Lett. {\bf 84} (2000) 4461.
\bibitem{rf:Fukumoto2} Y. Fukumoto: J. Phys. Soc. Jpn. {\bf 69} (2000) 2755.
\bibitem{rf:Knetter} C. Knetter, A. B\"{u}hler, E. M\"{u}ller-Hartmann and G. S. Uhrig: 
 Phys. Rev. Lett. {\bf 85} (2000) 3958.
\bibitem{rf:Nojiri} H. Nojiri, H. Kageyama, K. Onizuka, Y. Ueda and M. Motokawa: 
   J. Phys. Soc. Jpn. {\bf 68} (1999) 2906.
\bibitem{rf:Kageyama2} H. Kageyama, M. Nishi, N. Aso, K. Onizuka, 
   T. Yoshihama, K. Nukui, K. Kodama, K. Kakurai and Y. Ueda:
 Phys. Rev. Lett. {\bf 84} (2000) 5876.
\bibitem{rf:Lemmens} P. Lemmens, M. Grove, M. Fischer, G. G\"{u}ntherodt, V. N. Kotov,
   H. Kageyama, K. Onizuka and Y. Ueda: Phys. Rev. Lett. {\bf 85} (2000) 2605.
\bibitem{rf:Miyahara2} S. Miyahara and K. Ueda: Phys. Rev. B{\bf 61} (2000) 3417.
\bibitem{rf:Momoi} T. Momoi and K. Totsuka, Phys. Rev. B{\bf 61} (2000) 3231.
\bibitem{rf:Fukumoto} Y. Fukumoto and A. Oguchi: J. Phys. Soc. Jpn. {\bf 69} (2000) 1286.
\bibitem{rf:Momoi2} T. Momoi and K. Totsuka: cond-mat/0006020.
\bibitem{rf:Hartmann} E. M\"{u}ller-Hartmann, R. R. P. Singh, C. Knetter and G. S. Uhrig:
   Phys. Rev. Lett. {\bf 84} (2000) 1808.
\bibitem{rf:Onizuka} K. Onizuka, H. Kageyama, Y. Narumi,
   Y. Ueda and T. Goto: J. Phys. Soc. Jpn. {\bf 69} (2000) 1016.
\bibitem{rf:Tachiki} M. Tachiki and T. Yamada: 
   J. Phys. Soc. Jpn. {\bf 28} (1970) 1413.
\bibitem{rf:Fukumoto3} Y. Fukumoto and A. Oguchi: Proc. ICM 2000 (in press).
\end{thebibliography}
\end{document}